\documentclass[journal, 10pt]{IEEEtran}
\usepackage{graphicx,subfig}
\usepackage[cmex10]{amsmath}
\usepackage{algorithmic ,algorithm}
\usepackage{underscore}
\usepackage{url}
\usepackage{lipsum}
\usepackage{array}
\usepackage{mdwmath}
\usepackage{mdwtab}
\usepackage{multirow}
\usepackage{color}

\begin{document}
\title{A Qualitative Comparison of MPSoC Mobile and Embedded Virtualization Techniques}
\author{Junaid Shuja$^a$, Abdullah Gani$^a$, Sajjad A. Madani$^b$ \\ $^a$ Faculty of Computer Science and Information Technology, University of Malaya, Malaysia\\ $^b$ COMSATS Institute of Information Technology, Islamabad\\ Email: junaidshuja@siswa.um.edu.my, abdullah@um.edu.my, madani@comsats.edu.pk} 

\maketitle

\begin{abstract}
Virtualization is generally adopted in server and desktop environments to provide for fault tolerance, resource management, and energy efficiency. Virtualization enables parallel execution of multiple operating systems (OSs) while sharing the hardware resources. Virtualization was previously not deemed as feasible technology for mobile and embedded devices due to their limited processing and memory resource. However, the enterprises are advocating Bring Your Own Device (BYOD) applications that enable co-existence of heterogeneous OSs on a single mobile device. Moreover, embedded device require virtualization for logical isolation of secure and general purpose OSs on single device. In this paper we investigate the processor architectures in the mobile and embedded space while examining their formal virtualizabilty. We also compare the virtualization solutions enabling coexistence of multiple OSs in Multicore Processor System-on-Chip (MPSoC) mobile and embedded systems. We advocate that virtualization is necessary to manage resource in MPSoC designs and to enable BYOD, security, and logical isolation use cases.
\end{abstract}


\section{Introduction}\label{sec:intro}
Virtualization technology has been deployed in server space for decades since the early IBM mainframe systems~\cite{Popek1974}. In server space, virtualization enables hosting of multiple OSs on a single hardware platform. The hosted OSs are often homogeneous providing fault tolerance, resource management in terms of workload migration and consolidation, and energy efficiency. The Hypervisor or Virtual Machine Monitor (VMM) is placed between the hardware and host OSs and acts as a super OS. The hypervisor needs to efficiently share the hardware resources among guest OSs also known as Virtual Machines (VM). Moreover, hypervisor needs to translate guest OSs instructions if they do not meet the profile of underlying hardware. While translating guest OS instructions, virtualization adds the flexibility to execute non-compatible OS hardware combinations~\cite{Aguiar2010}.

Mobile and embedded devices are resource constrained and OSs are reduced to minimal lines of code before deployment. Hosting multiple OSs on such resource constrained devices seems performance prohibitive. Mobile and embedded device such as Electronic Control Units (ECUs) can lose hard real-time functionality due to overheads of virtualization. However, due to recent advances in mobile technology such as multicore processor designs and higher memory volumes, virtualization has become feasible. There are three main use cases of virtualization in mobile and embedded systems: \textbf{(a)} virtualization enables BYOD scenario where an employee can execute both enterprise and personal applications on a single mobile device, \textbf{(b)} virtualization enables isolation of mixed criticality OSs on a device, and \textbf{(c)} virtualization leads to energy efficient resource utilization~\cite{Shuja2012,Heiser2008,Shuja2014}. 

ARM based processors capture 90\% of smartphone market~\cite{Do2011}. However, there are a host of other processor architectures in the mobile and embedded space such as PowerPC~\cite{Mittal2013}, MIPS~\cite{Aguiar2011}, and SH-4A~\cite{Kanda2008} that are found in automobile controllers and consumer electronics. To implement virtualization efficiently, support from the underlying hardware platform is required such that: \textbf{(a)} the processor implements a hypervisor mode where the hypervisor can reside in a higher privilege level rather than occupying privilege level of the guest OS, and \textbf{(b)} each instruction that tries to change the context of hardware resources is privileged instruction~\cite{Penneman2013}. Major mobile and embedded processor architectures including ARM lack the aforementioned hardware support for virtualization. Therefore, the hypervisor complexity increases as it needs to take care of privilege levels and instructions of the guest OSs. 

\begin{figure*}
\centering
\includegraphics[width=12cm]{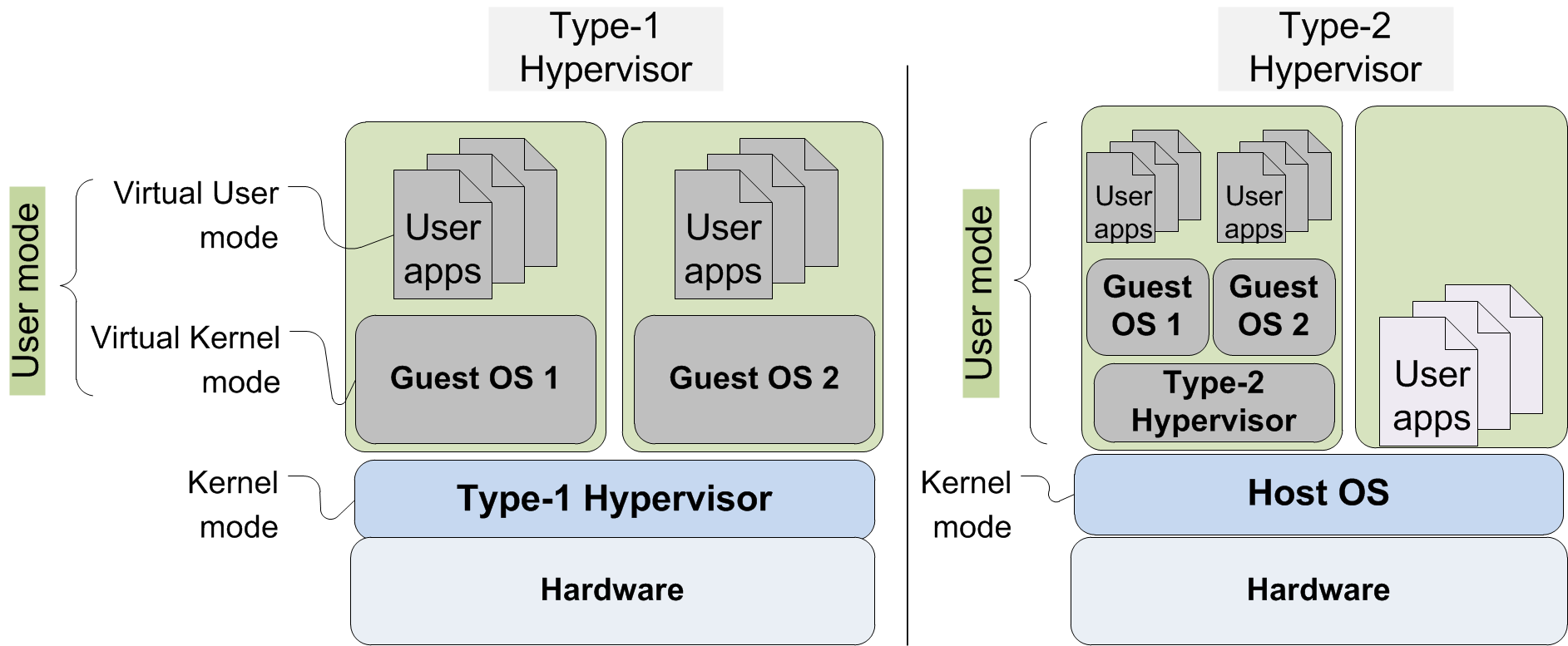}
\caption {Type-1 and Type-2 virtualization techniques}
\label{Fig:class}
\end{figure*}

Virtualization in mobile and embedded space is used to host heterogeneous OSs that address conflicting application requirements such as real time capability and user friendly GUI. Virtualization can enable concurrent hosting of a Real Time OS (RTOS) for sensor and communication control of the device and a general purpose OS (Android) for feature-rich GUI. IT enterprises usually control the technology their employees utilize during the office hours. This necessitates that each employee carry two mobile devices for enterprise and personal purposes. However, virtualization enables BYOD concept where both enterprise and personal OS and applications are hosted on a single mobile device in a logically isolated environment. Moreover, the BYOD concept significantly lowers the cost of hardware~\cite{Earley2014}. Embedded space devices often enforce strict security requirements on the hosted OS. However, virtualization enables concurrent execution of mixed criticality OSs such as a formally certified secure OS for automobile control and a general purpose OS for car entertainment system. Virtualization assures isolation of the domains by implementing separate address space for each hosted OS~\cite{Varanasi2011}. Furthermore, virtualization enables software reusability and survival of legacy software on newer platforms. It is estimated that most of smartphone devices are now shipped with a virtualization layer~\cite{GartnerGroup2008}. 

There a number of challenges pertaining to efficient virtualization of mobile and embedded devices. Most of devices from mobile and embedded space have limited memory size. The hypervisor needs to maintain record of address mapping and instruction translation for the guest OSs in the memory, thus, over-burdening the resource. Moreover, MPSoC designs are often heterogeneous and aggregate processors with different instruction sets. Each processor architecture in the heterogeneous MPSoC design has different requirements for virtualizability. Moreover, most of OSs developed for mobile and embedded devices are designed to execute on single core. Hence, MPSoC designs require greater adoption and complexity in hypervisor design~\cite{Aguiar2011a}. However, virtualization is necessary to manage resources in trending MPSoC architectures. Hardware support of virtualization in mobile and embedded devices is still in its early stages. Devices already deployed in smartphones and consumer electronics lack hardware support features for virtualization. Software enabled virtualization techniques utilized in such cases are not as efficient and consumes extra memory resources~\cite{Brash2010}.

In this article, we provide a comparative analysis of software and hardware enabled virtualization techniques in mobile and embedded devices with particular focus on virtualization solutions addressing MPSoC designs. In section~\ref{sec:back} we provide a taxonomy of software and hardware based mobile and embedded virtualization techniques. In section~\ref{sec:challenges} we address the challenges and issues to virtualization of MPSoC mobile and embedded device. Section~\ref{sec:hyp} provides detailed comparison of various MPSoC virtualization techniques present in mobile and embedded devices and lists research challenges and issues to MPSoC devices. Section~\ref{sec:conc} concludes the discussion with future directions.  

\section{Background}\label{sec:back}
In the forthcoming subsections, we describe the formal and informal requirements of virtualization that are necessary for understanding of taxonomy of virtualization techniques. Additionally, we provide a short taxonomy of mobile and embedded virtualization solutions.
\subsection{Formal and informal requirements of virtualization}\label{sec:form}
The instructions of an Instruction Set Architecture (ISA) can be categorized into three classes: \textbf{(a)} instruction that properly execute in privileged mode and trap in unprivileged mode are privileged instructions, \textbf{(b)} instructions that try to modify the context of the system state are sensitive instructions, and \textbf{(c)} all other instructions are innocuous. Based on categorization of processor instructions, there are three formal requirements for classic virtualizability of an ISA. Firstly, the interface provided by the hypervisor to guest OSs should be identical to that of the underlying hardware. Secondly, most of the guest OS instructions should directly execute over the hardware with a small subset of instructions trapping to the hypervisor. Lastly, the hypervisor should remain in control of all virtualized physical resources~\cite{Penneman2013}.

Aside from the aforementioned formal requirements of virtualization, the resource constrained mobile and embedded devices enforce informal requirements for efficient virtualization. Firstly, while hosting multiple OSs, the device should not compromise real time capability. For example, the automobile controller requires hard real time response in vehicle tracking and management systems. Secondly, the virtualization layer should be scalable to the number of guest OSs and underlying processor cores. Otherwise, the virtualization will add no performance benefit in comparison to a single core non-virtualized environment. Thirdly, virtualization should be non-intrusive and require minimum guest OS modifications. Fourthly, the hypervisor should provide formal proof of security and guest OS domain isolation~\cite{Heiser2011}. 
\begin{figure*}
\centering
\includegraphics[width=12cm]{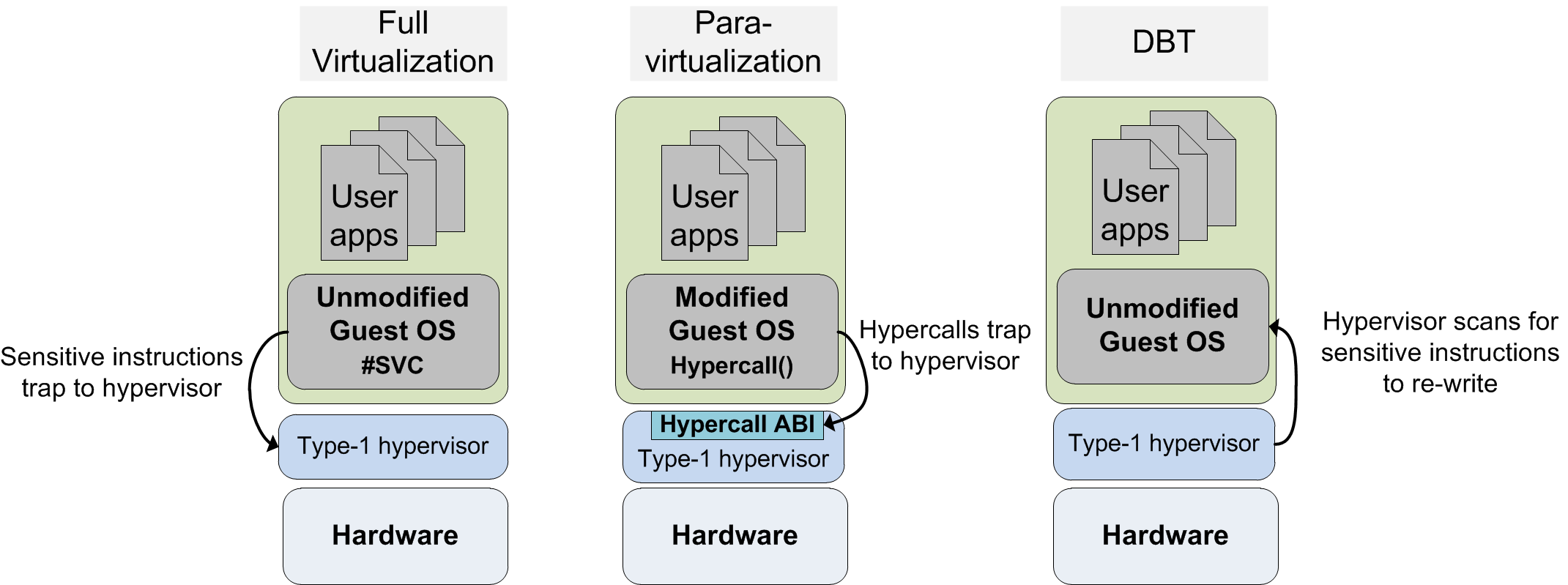}
\caption {Full, para, and DBT based virtualization techniques}
\label{Fig:type1}
\end{figure*}
\subsection{Virtualization techniques}\label{sec:virt}
Virtualization techniques are broadly categorized into Type-1 and Type-2 virtualization~\cite{Ding2012}. Type-1 virtualization enables the hypervisor to run in more privileged mode than the guest OSs. In this manner, the hypervisor fully controls access to the hardware resources made by guest OSs. Type-2 virtualization techniques execute the hypervisor as an application of the host OS. As a Type-2 hypervisor executes over an OS, it has less control of underlying hardware and can not provide resource sharing efficiently among hosted OSs. However, Type-2 hypervisors can be light-weight as they can host stripped down OSs environments~\cite{Barr2010}. Figure~\ref{Fig:type1} illustrates the difference between Type-1 and Type-2 virtualization techniques. 

Type-1 and Type-2 Virtualization techniques can be further classified into full virtualization, para virtualization, and Dynamic Binary Translation (DBT) techniques. In full virtualization, the hypervisor traps and emulates all instructions that try to change the context of hardware resources. The trap-and-emulate procedure is supported by either hardware or by DBT. Hardware support for full virtualization requires that the set of sensitive instructions is a subset of privileged instruction. In such scenario, all sensitive instructions will trap to the hypervisor for emulation. Full virtualization can also be implemented with the help of DBT by translation those sensitive instruction that do not trap at run time~\cite{Gu2012}. However, the instruction translation process introduces significant processing and memory overhead. To avoid translation of repeated set of instructions, the instructions with corresponding translation block are stored in memory for future reference. However, translation block storage transfers the overhead from processor to memory resources. Therefore, DBT based virtualization techniques are not suitable for mobile and embedded space devices. Paravirtualization requires the guest OS to be patched such that all sensitive non-privileged instructions are replaced by sensitive privileged instruction. Paravirtualization allows more hardware-OS platforms to be compatible. Instructions that are sensitive non-privileged are replaced by hypervisor calls. The hypervisor provides an interface to the guest OS for invocation of hypervisor calls. However, maintaining the patches of guest OS with upstream release versions increases the complexity of paravirtualization~\cite{Varanasi2010}. In case an ISA is not fully virtualizable, paravirtualization remains the only feasible solution. Lightweight paravirtualization avoids rigorous pre-patching of guest OS and translates the sensitive non-privileged instructions at runtime~\cite{Dall2010}. Microkernel based paravirtualization techniques are also popular in mobile and embedded devices~\cite{Heiser2010}. Microkernel based paravirtualization techniques emphasize the size reduction of the kernel to a minimal set while moving OS services from kernel space to user space. Figure~\ref{Fig:class} depicts full, para, and DBT based virtualization techniques.

\section{Mobile and embedded virtualization challenges}\label{sec:challenges}
Unfortunately, most of mobile and embedded ISAs do not meet the requirements of classic virtualizability~\cite{Suzuki2013}. ARM, PowerPC, and other processor ISA have introduced virtualization extensions to support efficient virtualization~\cite{Group2010a,Gilles2013}. However, the market release of processors with hardware support for virtualization is still in infancy. Most of processors powering current mobile and embedded devices lack hardware virtualization support. In this section we will discuss earlier ARM architectures as well as the hardware extensions and the challenges faced by both approaches.

Three major challenges to CPU virtualization in mobile and embedded devices are: \textbf{(a)} presence of sensitive non-privileged instructions in the ISA, \textbf{(b)} absence of hypervisor mode, and \textbf{(c)} heterogeneity of MPSoC architectures~\cite{Dall2014,Brash2010}. If an ISA contains sensitive non-privileged instructions, it is not classically virtualizable. Sensitive non-privileged instructions have to be replaced by either hypercalls through paravirtualization techniques or by DBT. While paravirtualization is not feasible for all OSs due to their closed source and propriety nature, DBT incurs memory overhead that might be prohibitive for resource constrained mobile and embedded devices. Some previous versions of architectures in mobile and embedded space such as ARM did not include a separate hypervisor mode. In such case, the hypervisor had to reside in the privileged OS mode, while the guest OS had to reside in the unprivileged user mode. As a result, all privileged instructions executed by the unprivileged guest OS trapped to the hypervisor for emulation. The trap-and-emulate process often degrades the real time performance of the hypervisor. Moreover, guest OSs had to reside in the same privilege level as their user applications. Resultantly, the hypervisor had to enforce domain access control mechanisms for isolation of guest OS and application memory space. Lastly, the MPSoC designs are currently trending in mobile and embedded space. Heterogeneous processors that offer different functionality are added to MPSoC design to support mixed criticality applications. As a result, virtualizability requirements of different processors add up, thus, increasing the hypervisor design complexity~\cite{Aguiar2011a}. 

The main challenges to sharing and virtualization of memory space in mobile and embedded devices are: \textbf{(a)} maintaining Shadow Page Tables (SPT) for multiple guest OSs, \textbf{(b)} memory space protection and isolation, and \textbf{(c)} Translation Lookaside Buffer (TLB) maintenance on context switches~\cite{Dall2013}. In a virtualized environment, the virtual address of the guest OS requires two level translation. Firstly, it is translated from guest virtual address to guest physical address by the guest OS. Then, guest physical address in translated to machine physical address by the hypervisor. The hypervisor has to maintain separate SPTs for each guest. Whenever, there is a change in guest OS page tables, the hypervisor has to capture and mirror that change in its SPT so that it is able to perform the second stage of address translation~\cite{Mijat2011}. Similarly, the hypervisor has to maintain some form of page table tagging mechanism so that it can isolate memory space of virtualized guest OSs. The hypervisor has to emulate domain tagging mechanism, if domain tagging support is not provided in hardware, such as old versions of ARM processors. Moreover, whenever a context switch takes place between the guest OSs or the hypervisor, TLB needs to be flushed and then re-populated for the currently executing host. As such context switches occur frequently in the virtualized environment, TLB misses increase resulting in high memory latency~\cite{Hwang2008}. 

Mobile and embedded devices consist of a plethora of sensors and communication boards such as WiFi, Bluetooth, and cellular radio. Sharing these components among multiple guest OSs requires fine grained timescale multiplexing. Moreover, GUI based devices also require consideration for GUI sharing and switching among multiple guest OSs. While the currently executed guest OS get full access of the GUI, the remaining guests have to wait for their turn~\cite{Barr2010}. Furthermore, special vector processors such as ARM Vector Floating-Point (VFP) requires virtualization support for resource sharing~\cite{Yoo2013}. 

\section{Comparison of mobile and embedded hypervisors}\label{sec:hyp}
There are a number of commercial and research based mobile and embedded hypervisors. We will focus on detailed comparison of those hypervisors that support MPSoC designs. 
\subsubsection{Virtual Hellfire Hypervisor}\label{sec:vhh} 
Virtual Hellfire Hypervisor~\cite{Aguiar2011a} is a MPSoC virtualization solution based on MIPS processors and Hellfire guest OSs. The Virtual Hellfire Hypervisor is microkernel based and provides a hypercall interface for the guest OSs to execute privileged instructions. The memory is partitioned into fixed chucks at kernel load time based on number of guests OSs. The intra-cluster communication between CPUs in the same cluster is done through hypercall interface while accessing a shared memory region among two clusters. Inter-cluster communication over the Network-on-Chip (NoC) architecture is done via wrapper functions. 
\begin{table*}
\centering
\caption{Comparison of MPSoC mobile hypervisors}
\label{tab:comp}    
\begin{tabular}{|m{2.6cm}|m{3.4cm}|m{1.2cm}|m{1.2cm}|m{1.8cm}|m{3.5cm}|m{1.2cm}|}
\hline
\textbf{Hypervisor} & \textbf{Technique} & \textbf{ISA} & \textbf{Real-time} & \textbf{Heterogeneous MPSoC} & \textbf{Overhead} & \textbf{Intrusive} \\
\hline
VHH~\cite{Aguiar2011a} & Microkernel based Paravirtual & MIPS & Yes & No & NA & Yes \\\hline
ITRI~\cite{Smirnov2013} & DBT and Paravirtual & ARM & No & No & 1.3-1.8 times slower for memory intensive apps & Yes \\\hline
KVM/ARM~\cite{Dall2014} & Full virtualization & ARM & No & No & within 10\% of native execution & No \\\hline
Proteus~\cite{Gilles2013} & Dual, DBT and Paravirtual & PowerPC  & Yes & No & $\geq$ 100 instruction cycles for virtualization related tasks & Yes \\\hline
AUTOSAR based~\cite{Reinhardt2014} & Paravirtual & AURIX & Yes & No & 40\% for memory intensive apps & Yes \\\hline
RAMPSocVM~\cite{Gohringer2011} & Paravirtual & PowerPC & Yes & No & NA & Yes \\\hline
\end{tabular}
\end{table*}

\subsubsection{ITRI}\label{sec:itri} 
ITRI~\cite{Smirnov2013} is an ARMv7 based hybrid hypervisor that is utilizes both paravirtualization and DBT based full virtualization. ITRI supports MPSoC design and multiple guest OSs in form of paravirtual Linux 2.6 kernels. The ITRI hypervisor is based on a modified version of Linux based Kernel Virtual Machine (KVM). The guest Linux kernel is patched by removing unnecessary privileged instructions and redundant context switches. The paravirtualization of Linux kernel helps reduce trap-and-emulate and TLB maintenance cost. QEMU emulator is used to run guest OS instances along with real and backend drivers. ITRI is one of few mobile hypervisors that supports guest migration~\cite{Groesbrink2014}. The performance evaluation of native and VM based execution show negligible overhead in compute intensive tasks while five times overhead in worst case for I/O intensive tasks.

\subsubsection{KVM/ARM}\label{sec:kvm} 
KVM/ARM~\cite{Dall2014} is light weight paravirtualization solution based on MPSoC ARM design. The light weight paravirtualization uses scripts based on regular expressions to replace sensitive non-privileged instructions with trap-and-emulate procedure. Inter-processor Interrupts (IPI) are used as a communication mechanism between multiple processors. In this context, hardware extensions for virtualizing the generic interrupt controller (GIC) are utilized. A virtual GIC interface is introduced for each CPU with a corresponding hypervisor control interface. Interrupt can be raised by writing to the list registers in the hypervisor control interface. Resultantly, the hypervisor directly raises hardware interrupt to the corresponding guest OS kernel without causing a trap. The performance evaluation shows 10\% overhead for realtime applications as compared to native execution.

\subsubsection{Proteus Hypervisor}\label{sec:prot} 
Proteus~\cite{Gilles2013} is a PowerPC architecture based hypervisor that supports MPSoC designs. Proteus is a hybrid hypervisor that implements both full and paravirtualization without specific hardware support. Although hardware support for virtualization was released in newer PowerPC devices, Proteus is based on previous releases. Proteus implements a microkernel by moving I/O device drivers to user space while the supervisor space only executes hypercall handler, VM scheduler, and inter-process communication modules.  Proteus design is Symmetric Multiprocessing (SMP); when a guest traps to the hypervisor, the hypervisor takes control of its assigned core. A small part of the hypervisor runs on each core to support full virtualization and forward inter-processor interrupts. Proteus hypervisor implements a semaphore based solution for access to shared devices in the user space. Experimental results show that virtualization overheads for most of instructions are below hundreds of processor cycles.    
 
\subsubsection{AUTOSAR based Hypervisor}\label{sec:auto} 
Reinhardt et al.~\cite{Reinhardt2014} proposed a hypervisor design for security critical automotive devices based on AURIX memory management less (MMU) processor and AUTOSAR guest OS. Automotive devices require domain and fault isolation between the guest OSs. Moreover, the number of guest OS and there memory requirements are fixed in such environments, allowing for fixed partitioning of memory and doing away with MMU. Instead Memory Protection Unit (MPU) is utilized for guest OS domain isolation. The hypervisor sets up itself and Virtual Device Emulators (VDEs) in the highest privileged mode while allocating accessible memory regions for each guest OS. The proposed hypervisor uses paravirtualization to emulate privileged instructions. Interrupts are first handled by the hypervisor and then forwarded to the corresponding guest OS through its vector table. The Microcontroller Abstraction Layer (MCAL) of AUTOSAR is ported to accommodate hypervisor startup code, guest OS memory allocations, and VDE setup for inter-VM communications. The paravirtualized solution results in large overheads over native executions.

\subsubsection{RAMPSoCVM}\label{sec:ramp} 
Runtime Adoptive Multiprocessor System-on-Chip Virtual Machine (RAMPSocVM)~\cite{Gohringer2011} utilizes embedded Linux kernel to extend support for MPSoC virtualization with Message Passing Interface (MPI). As the hypervisor is based on embedded Linux kernel, it comes with integrated support for multitasking and multithreading. As embedded Linux without BIOS detection, OpenFirmware is integrated as a module to detect dynamic MPSoC platform. The hypervisor manages the resources while meeting real-time constraints of the applications. To do so, CAP-OS is adopted as a module in the Linux kernel that services hard real-time constraints of applications. A MPI is provided to the applications to communicate with hypervisor services and device drivers managed by CAP-OS module. A Qualitative comparison of the discussed MPSoC mobile hypervisors is provided in Table~\ref{tab:comp}.

All the above discussed hypervisors have two main characteristics in common: \textbf{(a)} they are specific to mobile and embedded device space and based on such processor architectures, and \textbf{(b)} they are based on MPSoC architectures. However, despite of these studies, open research issues remain for future considerations. Most of the researchers have ignored the issue of heterogeneous MPSoC architectures, where different processor architectures enforce different requirements for virtualization. Although paravirtualization can enable virtualization of heterogeneous OSs, hardware supported full virtualization is desirable due to its efficiency. To the best of our knowledge, their is no such solution available in research and commercial virtualization techniques. Memory optimization for DBT based full virtualization is also not explored in the aforementioned virtualization solutions. While user responsiveness is driving mobile devices to higher power and enabling concurrent execution of multiple OSs, embedded devices still lack sufficient processing power and memory to support virtualization. Moreover, embedded devices often have hard real-time constraints that might be compromised in a resource sharing environment. Security and isolation of multiple domains can be addressed by formal verification of the hypervisor. However, such formal verifications are scarce in literature~\cite{Heiser2010a}.  

\section{Conclusion}\label{sec:conc}
In this paper, we carried out qualitative analysis of MPSoC design based mobile and embedded device virtualization solutions. We assert that the virtualization of mobile and embedded devices needs to be inspected from different aspects. As mobile devices such as smartphones and tablets are rapidly advancing in processor power and memory size, embedded devices are lacking behind. Most of the virtualization solutions are either paravirtual or DBT based full virtualization. Paravirtualization techniques require expensive guest OS patching while DBT based full virtualization solutions are inefficient in memory utilization. Hardware supported full virtualization is desired is such resource constrained platforms. Moreover, a performance analysis of paravirtualization, DBT based full virtualization, and hardware supported full virtualization techniques is yet to be debated. Furthermore, hardware supported full virtualization solutions for heterogeneous MPSoC designs need to be devised. 
\subsection*{Acknowledgments}
This work is partially funded by the Malaysian Ministry of Education under the High Impact Research Grant of University of Malaya UM.C/625/1/HIR/MOE/FCSIT/03.

\bibliographystyle{IEEEtran}
\bibliography{dc}
%

\end{document}